\documentclass[aps,prd,reprint,superscriptaddress]{revtex4-2}

\usepackage{amsmath,amssymb, amsthm,amstext}
\usepackage{natbib}
\usepackage{graphicx}
\usepackage{color}
\usepackage{array, enumerate}
\usepackage{bm}
\usepackage{multirow}
\usepackage[breaklinks,colorlinks,citecolor=blue]{hyperref}
\usepackage{braket}
\usepackage{txfonts}

\usepackage{fontawesome} 
\usepackage{orcidlink}
\usepackage{soul}

\graphicspath{{fig/}}

\newcommand{\lumdist}[0]{D_{\rm L}}
\newcommand{\dd}[0]{\mathrm{d}}

\newcommand{\secref}[1]{Sec.~\ref{#1}}
\newcommand{\figref}[1]{Fig.~\ref{#1}}
\newcommand{\equref}[1]{Eq.~(\ref{#1})}

\def\be{\begin{equation}}
\def\ee{\end{equation}}

\def\apjl{Astrophys.J.Lett.}

\def\jcap{JCAP}
\def\prd{Physical Review D}
\def\apj{The Astrophysical Journal}

\begin{document}
\title{Improving the detection sensitivity to primordial stochastic gravitational
waves with reduced astrophysical foregrounds. II. Subthreshold binary neutron stars}
\author{Mingzheng Li \orcidlink{0000-0002-3603-8532}}
\email{limz01@sjtu.edu.cn}
\affiliation{Tsung-Dao Lee Institute, Shanghai Jiao-Tong University, Shanghai, 520 Shengrong Road, 201210, People’s Republic of China}
\affiliation{School of Physics \& Astronomy, Shanghai Jiao-Tong University, Shanghai, 800 Dongchuan Road, 200240, People’s Republic of China}
\author{Jiming Yu \orcidlink{0000-0001-6319-0866}}
\email{jimingyu@sjtu.edu.cn}
\affiliation{School of Physics \& Astronomy, Shanghai Jiao-Tong University, Shanghai, 800 Dongchuan Road, 200240, People’s Republic of China}
\author{Zhen Pan \orcidlink{0000-0001-9608-009X}}
\email{zhpan@sjtu.edu.cn}
\affiliation{Tsung-Dao Lee Institute, Shanghai Jiao-Tong University, Shanghai, 520 Shengrong Road, 201210, People’s Republic of China}
\affiliation{School of Physics \& Astronomy, Shanghai Jiao-Tong University, Shanghai, 800 Dongchuan Road, 200240, People’s Republic of China}
\begin{abstract}
 Stochastic gravitational waves (GWs) consist of a primordial component from early Universe processes and an astrophysical component from compact binary mergers. To detect the primordial  stochastic GW background (SGWB), the astrophysical foregrounds must be reduced to high precision, which is achievable for third-generation (3G) ground based GW detectors. Previous studies have shown that the foreground from individually detectable merger events can be reduced with fractional residual energy density below $10^{-3}$, and the residual foreground from subthreshold binary neutron stars (BNSs) will be the bottleneck if not well cleaned. In this work, we propose that the foreground energy density of subthreshold BNSs $\Omega_{\rm sub}$ can be estimated via a population based approach from the individually detectable BNSs utilizing the isotropic orbital orientations of all BNSs, i.e., uniform distribution in $\cos\iota$, where $\iota$ is the BNS inclination angle with respect to the line of sight. Using this approach, we find $\Omega_{\rm sub}$ can be measured with percent-level uncertainty, assuming $O(10^5)$ individually detected BNSs in our simulations. This method represents a promising approach to tackling the foreground cleaning problem.
\end{abstract}
\maketitle

\section{Introduction} \label{sec:intro}

Primordial stochastic gravitational wave background (SGWB) is well motivated 
and has been speculated to be produced from various early Universe physical processes, including inflation \cite{Guth1981,Linde1982} and preheating \cite{Allahverdi2010,Amin2015}, first-order phase transitions \cite{Turner1990,Turner1992,Kamionkowski1994,Kosowsky1992,Caprini2008,Cutting2018} and cosmic strings \cite{Vilenkin1981,Hogan1984,Caldwell1992,Hindmarsh1995,Vilenkin2000,Buchmuller2021} (see \cite{Allen1997b,Chiara2016,Cai2017,Caprini2018,Christensen2019,Renzini2022} for complete reviews).
Primordial gravitational waves (GWs) have long been viewed as a unique probe to the Universe at the earliest moments. Therefore the primordial SGWB detection has been one of the primary targets for GW detectors in different frequency bands, including pulsar timing arrays \cite{Hobbs2010,McLaughlin2013,Hobbs2013,Lentati2015}, spaceborne GW detectors \cite{Thorpe2019,Mei2021} and ground based detectors \cite{AdvLIGO2015,AdvVirgo2015,KAGRA2013}.
And the first milestone is the recent detection of SGWB by pulsar timing arrays \cite{NANOGrav:2023gor,NANOGrav:2023hde,InternationalPulsarTimingArray:2023mzf, Reardon:2023gzh,Zic:2023gta, Antoniadis:2023rey,Antoniadis:2023utw, Xu:2023wog}, which has inspired intensive discussions about the implication to early universe processes \cite{NANOGrav:2023hvm,Antoniadis:2023xlr}, though it is still too early to attribute this detection to the primordial SGWB, due to the existing astrophysical foregrounds from supermassive black hole binaries \cite{NANOGrav:2023hfp, NANOGrav:2023pdq, Antoniadis:2023xlr, Bi:2023tib}.

Foreground cleaning is essential for detecting the primordial background in all frequency bands,
and in this work we will focus on the foreground cleaning problem of third-generation (3G) ground based detectors \cite{Regimbau2017,ET0000A18,Srivastava2022,ET2023}.
In a nutshell, this problem can be formulated as follows.
For an incoming binary merger event $d(t)$ at a detector, where $d(t) = h(t) + n(t)$ is the detector strain data, consisting of signal $h$ and noise $n$, one can estimate and subtract its contribution to the foreground energy density $\propto |h|^2$.
From the observable $d(t)$ and the detector noise power spectrum density (PSD) $P_{\rm n}(f)$,
one can construct various foreground cleaning methods, i.e., different ways of estimating and subtracting 
its contribution to the foreground energy density,
including the full Bayesian analysis where loud and subthreshold events are equally treated \cite{Drasco2003,Smith2019,Biscoveanu2020}, event identification and subtraction in the time-frequency domain \cite{Zhong2022},
and the classical cleaning method in the strain level via the subtraction of the maximum likelihood (ML) strain $h(\boldsymbol{\theta}^{\rm ML})$ where $\boldsymbol{\theta}^{\rm ML}$ represents the ML waveform model parameters \cite{Sharma2020},
and the optimization by further subtracting the expected value of the residue foreground energy density $\propto \braket{|\delta h(\boldsymbol{\theta}^{\rm ML})|^2}$ \cite{Pan2023} (hereafter Paper I). 
In addition to these methods, other background cleaning methods have been proposed in the literature,
applying to either signal strain $h(t)$ \cite{Cutler2006,Harms2008} or polarization modes $h_{+,\times}(t)$ \cite{Sachdev2020,Zhou2022,Zhou2022b,Song2024} instead of data $d(t)$ (see Sec.~III B of Paper I for detailed clarifications).

As shown in simulations of Paper I, the foreground from individually detectable merger events can be reduced with fractional residual energy density below $10^{-3}$, assuming a 3G GW detector network consisting of 2 Cosmic Explorers and 1 Einstein Telescope. Consequently, the residual foreground will be dominated by subthreshold binary neutron stars (BNSs), which has been a long-standing problem and has been recognized as the next critical problem to solve for detecting the primordial SGWB in the 3G era \cite{Zhu2013,Pan2023,Regimbau2017,Sachdev2020,Zhong2022,ET2023}. Both the classical cleaning method of subtracting the ML signal and the method of notching the individually resolved compact binary signals in time-frequency domain are incapable of cleaning subthreshold events, while this challenge may be addressed within a Bayesian framework, as discussed in \cite{Drasco2003,Smith2019,Biscoveanu2020}. One subtlety may be the enormous computational cost required in applying this method.

In this work, we propose a population based method for estimating the energy density of foreground from subthreshold BNSs. The basic idea is stated as follows. From the individually detected BNSs, we reconstruct their cosine inclination angle distribution, $p_{\rm det}(\cos\iota|\kappa)$, where $\kappa:=\mathcal{M}_z^{5/6}/\lumdist$ is the effective amplitiude with $\mathcal{M}_z$ representing the redshifted chirp mass and $\lumdist$ representing the luminosity distance.
From the distribution $p_{\rm det}(\cos\iota|\kappa)$ of detected BNSs, one can figure out the distribution of subthreshold BNSs $p_{\rm sub}(\cos\iota|\kappa)$ utilizing that the distribution of the whole population should be uniform in $\cos\iota$, i.e., $p_{\rm det}(\cos\iota|\kappa) + p_{\rm sub}(\cos\iota|\kappa) = \mathcal{U}(-1, 1)$. Then it is straightforward to calculate the energy density of foreground from subthreshold BNSs $\Omega_{\rm sub}$.

In practice, we perform a Bayesian population inference for the BNSs, constraining the population model $p_{\rm pop}(\cos\iota, \kappa|\mathbf{\Lambda})$ and the total number of all events $N_{\rm tot}$ from individually detected events $\{ \mathbf{d}_i\}$ $(i=1,...,N_{\rm det})$, where $N_{\rm det}$ is the number of the detections, and $\mathbf{\Lambda}$ denotes the BNS population model parameters. With the constrained population model $p_{\rm pop}(\cos\iota, \kappa|\mathbf{\Lambda})$ and the total number $N_{\rm tot}$ in hand, 
we can calculate the energy density of foreground from subthreshold BNSs and its uncertainty, $\Omega_{\rm sub}\pm \sigma (\Omega_{\rm sub})$. In fact, a similar idea based on compact binary population inference has been investigated and proven to be valuable for multiband foreground cleaning, reducing the mHz foreground of spaceborne GW detectors with 3G ground based detectors \cite{Pan:2019uyn}.

This paper is organized as follows. In \secref{sec:rev}, we give a brief review of foreground cleaning basics, emphasizing two approaches, an event-to-event approach and a population based approach, then introduce the hierarchical Bayesian method we use in the BNS population inference. In \secref{sec:clean_BNS} we explain the details in the Bayesian analysis of the BNS population, then present our results of constraining the BNS population model from a simulated sample of BNSs detected by the 3G detector network,
and the results of recovering GW foreground contributed by subthreshold BNSs. 
We summarize this paper in \secref{sec:Conclusion} with the conclusion that with proper cleaning, the astrophysical foregrounds of compact binaries may not be the limiting factor for detecting the primordial SGWB in the 3G era. 

In this paper, we use geometrical units $G=c=1$. We assume a flat $\Lambda$CDM cosmology with $H_0 = 67.7 ~{\rm km/s/Mpc}$, $\Omega_\Lambda = 0.69$ and $\Omega_{\rm m} = 0.31$, according to the Planck 2018 result \cite{planck18Cosmology}.

\section{Foreground cleaning basics}\label{sec:rev}

\subsection{A brief review}
In this work, we focus on cleaning the foreground from compact binaries, i.e., 
measuring and subtracting their contribution to the energy density of stochastic GWs.
Following the notations in Paper I, the energy density of stochastic GWs per logarithmic frequency is related to its power spectrum density (PSD) $H(f)$ by
\be\label{eq:omega}
\begin{aligned}
  \Omega_{\rm GW}(f)
  := \frac{1}{\rho_{\rm crit}} \frac{\dd\rho_{\rm GW}}{\dd \ln f} = \frac{4\pi^2}{3H_0^2}f^3 H(f)\ ,
\end{aligned}
\ee
where $\rho_{\rm crit}:=3 H_0^2/8\pi$ is the critical energy density to close the universe.
For the astrophysical foreground of compact binaries,
the PSD is formulated as (see e.g., \cite{Allen1999,Phinney2001,Pan:2019uyn} for derivation)
\be\label{eq:HA}
 H(f) = \frac{1}{T}\sum_i  \left(|h_+(f)|^2+|h_\times(f)|^2 \right)_i   \ ,
\ee
where the index $i$ runs over all binaries in the universe that merge within the observation time span $(0, T)$
, and $h_{+,\times}$ represent the two polarizations of incoming GWs. 
In terms of detector strain 
\be
\label{eq:strain}
  h(f) = F_+(\theta, \phi, \psi) h_+(f) + F_\times(\theta, \phi, \psi) h_\times(f)\ , \\
\ee
where the anttena pattern $F_{+,\times}$ depend on the source sky location ($\theta,\phi$) and the source polarization angle $\psi$, the PSD writes as
\begin{equation} \label{eq:HA_obs}
     H(f) = \frac{2}{\braket{F_+^2}+\braket{F_\times^2}}\frac{1}{T}\sum_i |h(f)|^2_i = \frac{5}{T}\sum_i |h(f)|^2_i\ ,
\end{equation}
where $\braket{}$ represents ensemble average over the three antenna pattern dependent angles,
and we have used the fact that $\braket{F_+^2}=\braket{F_\times^2} = 1/5$ for LIGO/Virgo/KAGRA (LVK) like L-shape interferometers in the 2nd equal sign \cite{Sathyaprakash2009}.
For a large population of compact binaries as we are investigating,
the foreground PSD can be calculated as 
\begin{equation} \label{eq:HA_sta}
    H(f) = \frac{5N_{\rm tot}}{T} \int |h(f;\boldsymbol{\theta})|^2 p_{\rm pop}(\boldsymbol{\theta}|\mathbf{\Lambda}) \dd \boldsymbol{\theta}\ ,
\end{equation}
where $N_{\rm tot}$ is the total number of the merger events during the observation time period $(0, T)$ and 
 $p_{\rm pop}(\boldsymbol{\theta}|\mathbf{\Lambda})$ is the population model with parameters $\mathbf{\Lambda}$, i.e., 
 the probability density of waveform model parameters $\boldsymbol{\theta}$, normalized as 
\be \label{eq:norm}
\int p_{\rm pop}(\boldsymbol{\theta} | \mathbf{\Lambda}) \dd \boldsymbol{\theta}= 1\ .
\ee 

The foreground cleaning problem is about measuring the foreground energy density or equivalently PSD,
and Eqs.~(\ref{eq:HA_obs},\ref{eq:HA_sta}) display two different perspectives in understanding the
astrophysical foreground PSD. The former represents an event-to-event approach and the latter represents 
a population based approach \cite{Pan:2019uyn}. Consequently, the foreground cleaning can be done following these two 
different approaches: in general, the former applies to loud events that are individually detectable,
while the latter better applies to a large population of events, especially a mixture of loud and subthreshold events.

\emph{Approach 1:}  For an incoming binary merger event, the detector strain data $d(f) = h(f) + n(f)$ consists of signal $h$ and noise $n$, the foreground cleaning is to estimate and subtract its contribution to the foreground energy density, or equivalently the strain magnitude $|h(f)|^2$ [see \equref{eq:HA_obs}] from data $d(f)$ and the detector noise power spectrum density $P_{\rm n}(f)$.
Note that neither the signal strain $h(f)$ nor the two polarization modes $h_{+,\times}(f)$ is the directly measured detector data strain.
There have been some unnecessary debates in the literature arising from applying the cleaning methods to these nondata and unknown quantities instead of the data $d(f)$.

For individually detectable events with signal to noise ratio (SNR) above
the detection threshold $\rho > \rho_{\rm thr}$, a foreground cleaning method 
has been detailed in Paper I and we briefly review as follows.
From data $d(f)$ and the detector noise power spectrum $P_{\rm n}(f)$, one can infer the maximum likelihood (ML) signal $h(f; \boldsymbol{\theta}^{\rm ML})$,
where $ \boldsymbol{\theta}^{\rm ML}$ is the ML/best-fit waveform model parameters.
As a first step, one can subtract the ML signal from data and the residual data is 
\be 
\delta d = d-h^{\rm ML} = (h-h^{\rm ML}) + n= \delta h + n\ .
\ee 
It is straightforward to find out that the residual power scales as $|\delta h|^2/|h|^2 \sim \rho^{-2}$, i.e., 
$|\delta h|^2\sim \rho^{0}$.
After the first step, there is no way to further clean the residual strain $\delta h$, since this will require a better estimate of the signal strain than the ML/best-fit estimate $h^{\rm ML}$.
But the expected value of the residual power  $\braket{|\delta h(f;\boldsymbol{\theta}^{\rm true})|^2}$ can be computed if the true model parameters $\boldsymbol{\theta}^{\rm true}$ were known.
Without the knowledge of the true parameters, an approximate estimator can be constructed using the ML parameters $\boldsymbol{\theta}^{\rm ML}$ as $\braket{|\delta h(f;\boldsymbol{\theta}^{\rm ML})|^2}$.
As shown in Paper I, after subtracting the approximate average residual power, the residue
is further reduced with fractional uncertainty scaling as $\left(|\delta h|^2 - \braket{|\delta h(\boldsymbol{\theta}^{\rm ML})|}\right)/|h|^2 \sim \rho ^{-3}$.

\emph{Approach 2:} The event-to-event approach above does not apply to subthreshold events since they are individually indistinguishable from detector noise fluctuations.  For a 3G detector network, almost all the BBHs are individually detectable while nearly half of BNSs are subthreshold, which has long been identified as a bottleneck for detecting the primordial SGWB in the 3G era \cite{Zhu2013,Pan2023,Regimbau2017,Sachdev2020,Zhong2022,ET2023}.
In this work, we will focus on cleaning the foreground from subthreshold BNSs in the population based approach.
The basic idea is rather straightforward. 
Using the fact that the orientations of all BNSs should be statistically isotropic, 
i.e., the distribution of $\cos\iota$ is uniform in the range of $[-1,1]$,
$p_{\rm det}(\cos\iota|\kappa) + p_{\rm sub}(\cos\iota|\kappa) = \mathcal{U}(-1, 1)$,
with $\iota$ representing the inclination between the BNS orbital angular momentum direction and the line of sight direction, 
one can  infer the number of subthreshold BNSs from the number of loud BNSs.
Then it is straightforward to calculate the energy density of foreground from subthreshold BNSs $\Omega_{\rm sub}$.
In practice, the above analysis should be conducted in the framework of Bayesian population inference,
constraining the population model $p_{\rm pop}(\cos\iota, \kappa|\mathbf{\Lambda})$ from individually detected events $\{ \mathbf{d}_i\}$ $(i=1,...,N_{\rm det})$, then calculating 
the energy density (and its uncertainty) of foreground from subthreshold BNSs 
$\Omega_{\rm sub}\pm \sigma (\Omega_{\rm sub})$.

\subsection{Bayesian population inference}

In this subsection, we will explain the basics of Bayesian population inference, starting with the model parameter inference of an individual GW event.

For a network of $N_\mathrm{d}$ GW detectors, the strain data can be written as
\begin{equation}
    \mathbf{d}(f)=\left[d_1(f),\cdots,d_{N_d}(f)\right]^T e^{-i\mathbf{\Phi}},
\end{equation}
where $\mathbf{\Phi}$ is a diagonal matrix with
\begin{equation}
    \Phi_{IJ}=2\pi f\delta_{IJ}\tau_I,
\end{equation}
which represents the time delay for GW signals to reach each detector. In this work, we adopt the IMRPhenomD waveform model \cite{husa2016Frequencydomain, khan2016Frequencydomain} and use model parameters $\boldsymbol{\theta}=(\theta, \phi, \psi, \iota, \kappa, \mathcal{M}_z, q,  t_c, \psi_c)$, where $\theta, \phi, \psi$ are the direction angles
and the polarization angle of the source, $\kappa:=\mathcal{M}_z^{5/6}/\lumdist$ is the effective amplitude as introduced in \secref{sec:intro}, $t_c$ and $\psi_c$ are the coalescence time and phase, respectively. 
In principle, the effects of tidal deformation and star spins should also be taken in account in realistic data analysis, here we neglect these minor effects for convenience and saving the computational time, as they are not expected to largely change the forecast results. 

From the Bayes' theorem, the posterior of parameters $\boldsymbol{\theta}$ constrained by data $\mathbf{d}$  is formulated as 
\begin{equation}
    \mathcal{P}(\boldsymbol{\theta}|\mathbf{d})\propto \mathcal{L}(\mathbf{d}|\boldsymbol{\theta})\pi(\boldsymbol{\theta}),
\end{equation}
where $\mathcal{L}(\mathbf{d}|\boldsymbol{\theta})$ is the likelihood of detecting data $\mathbf{d}$
in the detector network for an incoming GW signal parametrized by $\boldsymbol{\theta}$, and $\pi(\boldsymbol{\theta})$ is the prior of the parameters assumed. In GW data analysis, the likelihood $\mathcal{L}(\mathbf{d} | \boldsymbol{\theta})$ is defined as \cite{abbott2021GWTC2}
\begin{equation}
    \label{eq:EventLikelihood}
    \mathcal{L}(\mathbf{d} | \boldsymbol{\theta}) \propto \exp\left\{ \sum_{I=1}^{N_\mathrm{d}}\left[-\frac{1}{2} \left\langle d_I-h(\boldsymbol{\theta}) \mid d_I-h(\boldsymbol{\theta}) \right\rangle \right]\right\},
\end{equation}
where the noise weighted inner product $\langle a|b \rangle$ is defined as 
\begin{equation}
    \label{eq:InnerProductDef}
    \left\langle a | b \right\rangle \equiv 4 \mathrm{Re} \int_{f_{\rm min}}^{f_{\rm max}} \frac{a(f) b^{*}(f)}{P_{\rm n}(f)} \dd f,
\end{equation}
with $P_{\rm n}(f)$  being the detector noise PSD.
Following the discussions in Ref.~\cite{Borhanian2021}, we consider a reference detector network consisting of a 40 km Cosmic Explorer, a 20 km Cosmic Explorer and a Einstein Telescope (see Fig.~1 in Paper I for a visual summary of the detector noise PSDs). 

From loud events that can be individually detected $\{ \mathbf{d}_i\} \ (i=1,...,N_{\rm det})$, one can infer the total number of all events $N_{\rm tot}$ and the population parameters $\mathbf{\Lambda}$ using the hierarchical Bayesian method, with the population likelihood \cite{LVCO1O2,LVKCO3}

\begin{equation}
    \label{eq:PopLikelihood}
    \mathcal{L}(\{ \mathbf{d}_i\} | \mathbf{\Lambda}, N_{\rm tot}) \propto N_{\rm tot}^{N_{\rm det}} e^{-N_{\rm tot} \xi(\mathbf{\Lambda})} \prod_{i=1}^{N_{\rm det}} \int \mathcal{L}\left( \mathbf{d}_i | \boldsymbol{\theta}\right) p_{\rm pop}(\boldsymbol{\theta} | \mathbf{\Lambda}) \dd \boldsymbol{\theta} \ .
\end{equation}
The $\xi(\mathbf{\Lambda})$ term represents the fraction of detectable BNSs in the population $p_{\rm pop}(\boldsymbol{\theta} | \mathbf{\Lambda})$
and is defined as 
\begin{equation}
    \label{eq:XiDef}
    \xi(\mathbf{\Lambda})  = \int \dd\boldsymbol{\theta} ~ \Theta(\rho_{\rm obs}(\boldsymbol{\theta})-\rho_{\rm thr}) p_{\rm pop}(\boldsymbol{\theta} | \mathbf{\Lambda}),
\end{equation}
where $\Theta$ is the Heaviside step function, i.e., only loud events with observed SNR above the detection threshold $\rho_{\rm obs} > \rho_{\rm thr}$ are classified as detected. Note that the ML parameters $\boldsymbol{\theta}^{\rm ML}$ inferred from data differ from the true parameters $\boldsymbol{\theta}^{\rm true}$ due to detector noises, consequently the observed SNR $\rho_{\rm obs} := \rho(\boldsymbol{\theta}^{\rm ML})$ is not equal to the true SNR $\rho_{\rm true}:=\rho(\boldsymbol{\theta}^{\rm true})$. Instead, $\rho_{\rm obs}$ fluctuates around the true value with a unit standard deviation, i.e., 
\begin{equation}
    \label{eq:SNRObsDistribution}
    \rho_{\rm obs} \sim \mathcal{N}(\rho_{\rm true}, 1)\ ,
\end{equation}
which makes a nontrivial difference and therefore is necessary to be considered in the calculation of $\xi(\mathbf{\Lambda})$. As long as $\xi(\mathbf{\Lambda})$ and then the population likelihood are calculated [\equref{eq:PopLikelihood}], the BNS population model $p_{\rm pop}(\boldsymbol{\theta}|\mathbf{\Lambda})$ and the total number $N_{\rm tot}$ can be constrained, and the foreground PSD of BNSs can therefore be calculated using \equref{eq:HA_sta}.

\section{Cleaning the foreground of subthreshold BNS}\label{sec:clean_BNS}

As explained in the previous section, the key step of cleaning the foreground of subthreshold BNSs is the BNS population inference [Eq.~(\ref{eq:PopLikelihood})].  The commonly used method of evaluating the high-dimensional integrals is replacing the integrals with Monte Carlo estimations \cite{LVKCO3,gray2020Cosmological}, 
\begin{equation}\label{eq:popinfer}
    \int \mathcal{L}\left( \mathbf{d}_i | \boldsymbol{\theta}\right) p_{\rm pop}(\boldsymbol{\theta} | \mathbf{\Lambda}) \dd \boldsymbol{\theta}
    \approx \frac{1}{N_{\rm MC}} \sum_{
        \substack{\boldsymbol{\theta}_j\sim \mathcal{P}(\boldsymbol{\theta}| \mathbf{d}_i) \\ j=1, \cdots, N_{\rm MC}}}
    \frac{p_{\rm pop}(\boldsymbol{\theta}_j | \mathbf{\Lambda})}{\pi(\boldsymbol{\theta}_j)}\ ,
\end{equation} 
where $N_{\rm MC}$ is the total number of sampled parameter sets in the Bayesian parameter inference of data $ \mathbf{d}_i$.
The computation intensive part is the Bayesian parameter inference $\mathcal{P}(\boldsymbol{\theta}| \mathbf{d}_i)$ for each event. 
In practice, this part will be done by the Cosmic Explorer and Einstein Telescope collaborations as what the LIGO/Virgo/KAGRA collaboration has done.
With the parameter samples $\boldsymbol{\theta}_j\sim \mathcal{P}(\boldsymbol{\theta}| \mathbf{d}_i)$ provided, 
implementing Eq.~(\ref{eq:popinfer}) is not computationally intense.
In this work, there is no such samples provided for free of course. 
It seems that we have to generate such parameter samples via a large number of MCMC simulations on mock data with signal injections.
Actually we find this is not necessary for the purpose of sensitivity forecast.
The reason is that only a subset of waveform parameters matters and implementing Eq.~(\ref{eq:popinfer}) can be simplified.

\subsection{BNS population inference}
In the sensitivity band of 3G detectors, the amplitude of an inspiralling BNS can be well approximated as  \cite{husa2016Frequencydomain, khan2016Frequencydomain} 
\be 
|h(f;\boldsymbol{\theta})| \propto \kappa f^{-7/6}
\sqrt{F_+^2 \left(\frac{1+\cos^2\iota}{2}\right)^2 + F_\times ^2 \cos^2\iota}\ ,
\ee 
where  $F_{+,\times}(\theta, \phi, \psi)$ are the antenna pattern functions.
For the purpose of evaluating the foreground PSD of BNSs, we only need the distribution of a subset of waveform parameters
$\{\kappa, \iota, \theta, \phi, \psi\}$.
In any reasonable population model $p_{\rm pop}(\boldsymbol{\theta} | \mathbf{\Lambda})$, 
the distributions of angles $\{\iota, \theta, \phi, \psi\}$ 
are determined by the isotropy of the universe and are naturally known.
Therefore only the amplitude distribution $p_\kappa(\kappa)$ remains to be determined from the detected events.
For any BNS merger event, the chirp mass $\mathcal{M}_z$ is most tightly constrained among all the waveform parameters,
while the constraint on $\lumdist$ and therefore on $\kappa$ is limited by the degeneracy with the inclination angle $\iota$.
This is the well known $\lumdist-\iota$ degeneracy. In addition, the chirp mass $\mathcal{M}_z$,
the source direction $\hat{n}$ and $(\kappa, \iota)$ are mainly constrained 
by the phase/frequency evolution (chirp),  arrival time differences to different detectors \cite{wen:2010Geometricala}
and the signal strength, respectively. Therefore, the uncertainties of $(\kappa, \iota)$ are dominated by their mutual correlation,
and their correlations with other parameters are weak (see  Ref.~\cite{roulet2022Removing} for a comprehensive investigation of 
removing degeneracy with reparametrization).
Based on this observation, we divide the waveform parameters into two subsets, $\boldsymbol{\theta}=\{\kappa, \cos\iota\} \oplus \boldsymbol{\theta}_1$, write the population model as $p_{\rm pop}(\boldsymbol{\theta}|\mathbf{\Lambda}) =p_\kappa(\kappa) p_1(\boldsymbol{\theta_1}|\mathbf{\Lambda_1})$, and rewrite the integral in \equref{eq:PopLikelihood} as 
\be 
    \int \mathcal{L}\left( \mathbf{d}_i | \boldsymbol{\theta}\right) p_{\rm pop}(\boldsymbol{\theta} | \mathbf{\Lambda}) \dd \boldsymbol{\theta} \\ 
    = \int \mathcal{L}\left( \mathbf{d}_i | \kappa, \cos\iota \right) p_{\kappa}(\kappa) ~\dd \kappa ~\dd \cos\iota\ ,
\ee 
where
\be\label{eq:like2d} 
\mathcal{L}\left(  \mathbf{d}_i|\kappa, \cos\iota  \right) = \int \mathcal{L}( \mathbf{d}_i|\boldsymbol{\theta}) p_1(\boldsymbol{\theta}_1 | \mathbf{\Lambda}_1) ~\dd\boldsymbol{\theta_1}\ ,
\ee 
is the two-dimensional likelihood marginalized over all other waveform parameters $\boldsymbol{\theta}_1$.

One method of evaluating $\mathcal{L}\left( \mathbf{d}_i | \kappa,\cos\iota \right)$ is calculating the Fisher matrix $F$ for each event, and the likelihood can be approximated as a Gaussian distribution centered on the true parameters and 
with covariance matrix $F^{-1}$. There are two potential issues in this approximation.
One is that the parameter uncertainties and correlations inferred from Fisher matrix are only valid for high-SNR events \cite{vallisneri2008Use}, while events with SNR near the threshold $\rho_{\rm thr}$ provide essential information for the population inference. The other is that the true parameters $\boldsymbol{\theta}^{\rm true}$ are unknown in practice, therefore the population inference result obtained from the likelihood given by Fisher approximation could be biased from the inference result of real-world observations.

\begin{figure}
    \includegraphics[width=\columnwidth]{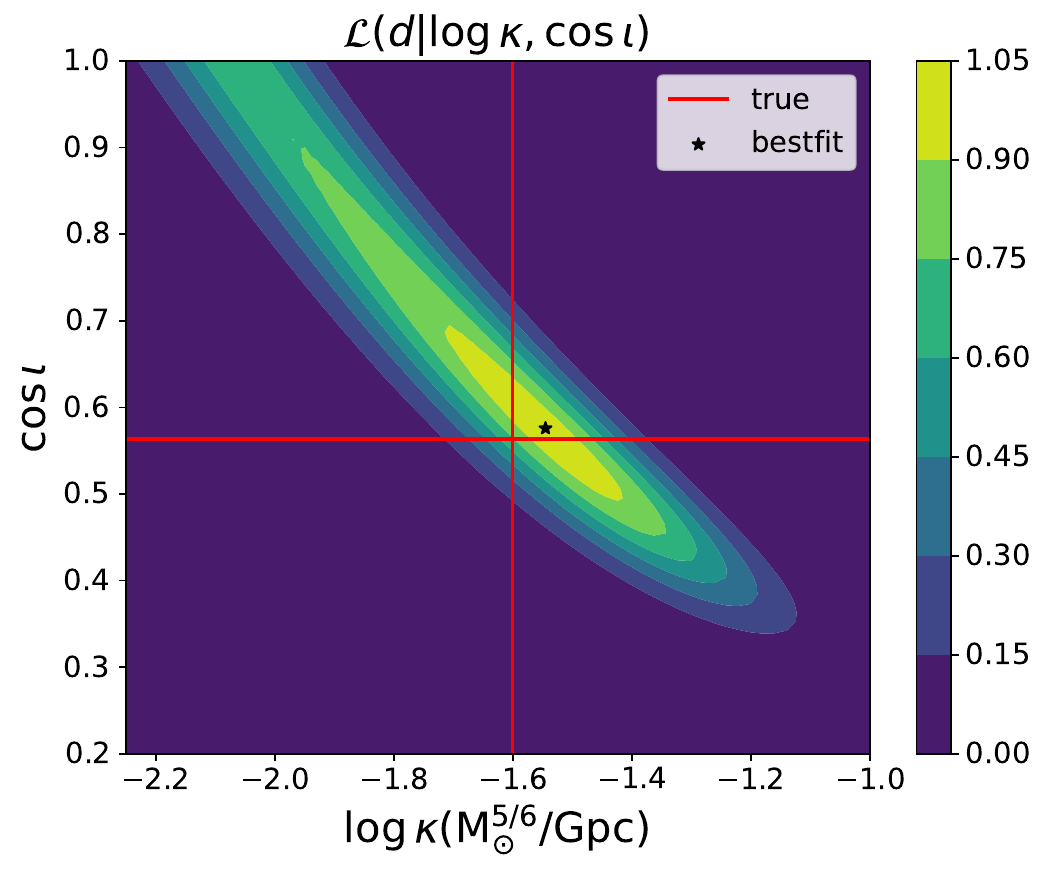}
    \caption{The likelihood $\mathcal{L}( \mathbf{d}_i | \log\kappa,\cos\iota)$ for an example event. Vertical and horizontal lines denote the true values of $\log \kappa$ and $\cos\iota$, while the black star denotes the best fit value. The likelihood is normalized to make sure the maximum value being unity, in order to avoid numerical errors on extremely small numbers.}
    \label{fig:KappaLikelihood2D}
\end{figure}

\begin{figure}
    \includegraphics[width=\columnwidth]{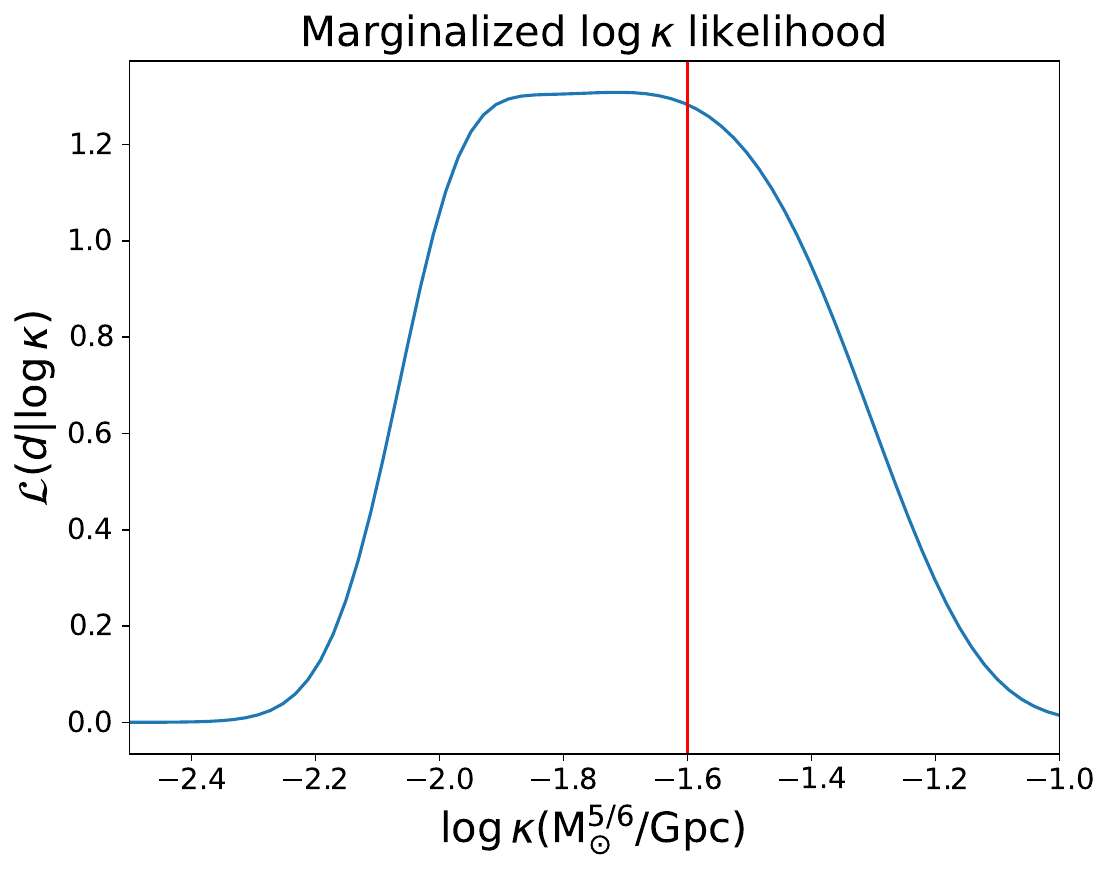}
    \caption{Marginalized likelihood $\mathcal{L}( \mathbf{d}_i | \kappa)$ for the same event as \figref{fig:KappaLikelihood2D}. The red vertical line denotes the true value of $\log \kappa$.}
    \label{fig:KappaLikelihood1D}
\end{figure}

To avoid these two issues and mimic the realistic data analysis more closely, we choose to directly calculate the likelihood $\mathcal{L}(\mathbf{d} | \kappa, \cos\iota)$ using \equref{eq:EventLikelihood}: we sample the Gaussian noise $n(f)$ using the power spectrum $P_{\rm n}(f)$ of each detector to obtain the detector strain $d(f)=h(f)+n(f)$. 
Using the approximation that the uncertainties of $\kappa$ and $\cos\iota$ are mainly due to their mutual correlation, while their correlations with other parameters $\boldsymbol{\theta}_1$ are negligible, 
the evaluation of Eq.~(\ref{eq:like2d}) 
is approximated as 
\be 
\mathcal{L}(\mathbf{d} | \kappa,\cos\iota) 
= \int \mathcal{L}( \mathbf{d}_i|\boldsymbol{\theta}) p_1(\boldsymbol{\theta}_1 | \mathbf{\Lambda}_1) ~\dd\boldsymbol{\theta_1}
\propto \mathcal{L}(\mathbf{d} | \kappa,\cos\iota,\boldsymbol{\theta}_1^{\rm true})\ .
\ee 
For each event, we calculate the likelihood $\mathcal{L}( \mathbf{d}_i | \kappa,\cos\iota)$ on $100\times 100$ linear grid of $\log \kappa$ and $\cos\iota$, with $\cos\iota$ ranging from $[-1,1]$ and $\log \kappa$ ranging from full possible range given by mass and redshift distribution. \figref{fig:KappaLikelihood2D} displays the likelihood $\mathcal{L}( \mathbf{d}_i | \log\kappa,\cos\iota)$ for an example event, which clearly shows the strong degeneracy between $\kappa$ and $\iota$. The marginalized likelihood $\mathcal{L}( \mathbf{d}_i | \log\kappa) :=\int \mathcal{L}( \mathbf{d}_i | \log\kappa,\cos\iota) \ \dd \cos\iota$ for the same event is shown in \figref{fig:KappaLikelihood1D}, which displays more features than the Gaussian likelihood given by the simple Fisher matrix approximation.

To summarize, the steps of inferring the BNS population model from a number of simulated BNS events are stated as follows:
(1) generating of BNS events with total number $N_{\rm tot}$, and parameters $\{\boldsymbol{\theta}\}_{i=1,...,N_{\rm tot}}$ sampled from an injection population model $p_{\rm pop}^{\rm inj}(\boldsymbol{\theta}|\mathbf{\Lambda})$;
(2) calculating the observed SNR $\rho_{\rm obs}$ of each event with \equref{eq:SNRObsDistribution}, and labeling loud events with $\rho_{\rm obs}\ge\rho_{\rm thr}$ as detected; 
(3) picking a population model $p_{\rm pop}(\boldsymbol{\theta}|\mathbf{\Lambda})$, constraining the population model $p_{\rm pop}(\boldsymbol{\theta}|\mathbf{\Lambda})$ and the total number $N_{\rm tot}$ of events in the framework of hierarchical Bayesian analysis 
with the population likelihood in \equref{eq:PopLikelihood}, 
the evaluation of which makes use of the  likelihood of individual events in \equref{eq:EventLikelihood}.
The implementation of these steps is coded in a \texttt{Python} package \texttt{PoppinGW} we developed, where \texttt{GWFAST} \cite{iacovelliTextttGWFASTFisher2022} is used in waveform related calculations and \texttt{BILBY} \cite{ashton2019BILBY} is used for conducting Bayesian inferences. Source code of \texttt{PoppinGW} is available on GitHub \faGithub~ \url{https://github.com/mzLi01/poppingw}.

\subsection{Examples of BNS population inference} \label{sec:RedshiftDistribution}

In this subsection, we aim to examine how accurately the BNS population model and therefore the foreground energy density of BNSs can be constrained with the hierarchical Bayesian method from simulated BNS observations with 3G detectors. 

As a fiducial population model, we use the same BNS merger rate model as in Paper I, 
\begin{equation}
    \label{eq:MergerRateReal}
    R(z)=R_0 (1+z)^{2.9} e^{-z^2/3} \quad {\rm for} \quad (z\le 6),
\end{equation}
with the local merger rate $R_0= 160 ~{\rm Gpc^{-3} yr^{-1}}$ \cite{LVCO1O2,LVKCO3}
(see e.g., \cite{Perigois2021,Zhou2022,Regimbau2022}
for more detailed rate modeling). Note that the choice of the local merger rate $R_0$ does not affect the results of population inference, as only the total number of BNS merger events $N_{\rm tot}$ matters in the inference.
The total merger rate in the observer frame is then
\begin{equation}
    \label{eq:MergerRateObs}
    \dot{N} = \int \frac{\dd V_c(z)}{\dd z} \frac{R(z)}{1+z} ~\dd z \ ,
\end{equation}
where $ V_c(z)$ is the comoving volume of the Universe within redshift $z$.
In the fiducial model, the total merger rate turns out to be $\dot{N}=3.7\times 10^5 {\rm yr^{-1}}$. 
For other parameters, we set 
\begin{equation}\label{eq:pop_bns}
\begin{aligned}
    &m_1,m_2\sim\mathcal{U}[1.2,2.5]\ M_\odot \ , \\ 
    & \cos\theta,\cos\iota\sim\mathcal{U}[-1,1]\ , \\  
    & \phi, \psi, \psi_c\sim\mathcal{U}[0,2\pi) \ ,
\end{aligned}    
\end{equation}
where $m_1, m_2$ are masses of the binary.

As a convenient example, we first sample a total number of $N_{\rm tot}=10^3\  (\approx \dot N\times 1 \ {\rm day})$ BNS events from the fiducial population model, $N_{\rm det} = 683$ events among these have SNR exceed the detection threshold, i.e., $\rho_{\rm obs}>\rho_{\rm thr}=10$. The population inference is performed on these individually detected events. To study the effect of accumulating observation time on the accuracy of the population inference, we also sample another set of events with $N_{\rm tot}=10^4$ from the same population, $N_{\rm det}=6939$ events among which are individually detected.

\begin{figure*}
    \includegraphics[width=\textwidth]{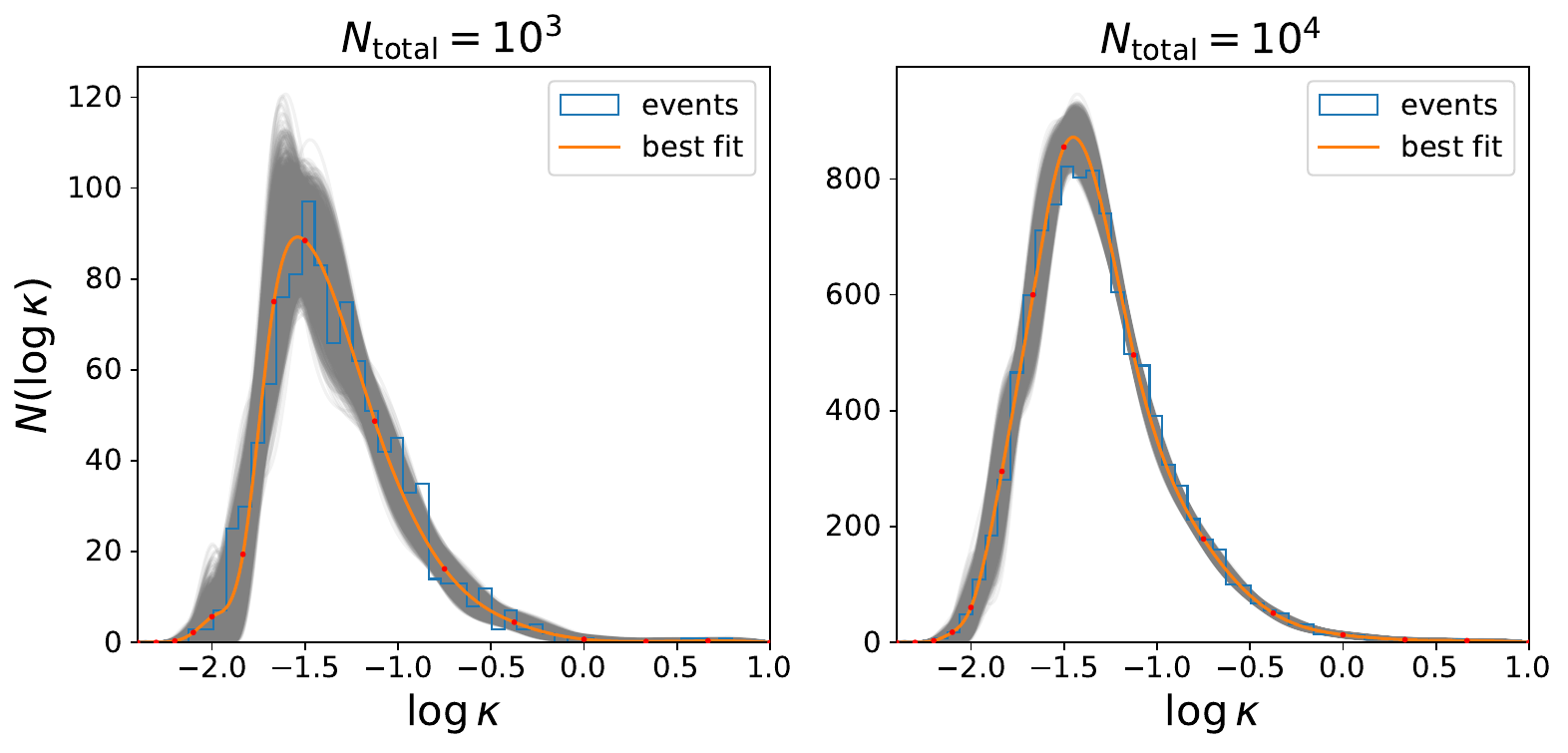}
    \caption{
    Constraint on the population model $p(\log\kappa)$ from the simulated $N_{\rm det}$ BNS mergers that are individually detected. 
    The left and right panel show the results given by $10^3$ and $10^4$ events individually. The orange line shows the best fit distribution, and the blue line is the histogram of the injected events.
    The gray lines denote the $1\sigma$ uncertainty range of $N(\log\kappa)$ from the population inference.}
    \label{fig:InferPkSpline}
\end{figure*}

Before implementing the Bayesian population inference with likelihood in Eq.~(\ref{eq:PopLikelihood}),
we need to determine the parametrization form of the population $p_\kappa(\log\kappa;\mathbf{\Lambda})$. To be as general as possible, we parametrize $p_\kappa(\log\kappa)$ using cubic spline interpolation among a number of discrete points $p(\log\kappa_i)$ with $i\in(0,\dots,14)$. The boundary values $p(\log\kappa_0)$ and $p(\log\kappa_{14})$ are fixed to be zero, considering the low probability of BNS mergers at extremely low and extremely high redshifts. 
In this general BNS population model with model parameters $\mathbf{\Lambda}=\{p_i\} ~(i=1,\dots,13)$, the probability density of given $\{p(\log\kappa_i)\}$ is expressed as
\begin{equation*}
    p(\log\kappa) = \frac{{\rm CubicSpline}(\log\kappa;\{p(\log\kappa_i)\})}{\int_{\log\kappa_{\rm min}}^{\log\kappa_{\rm max}} {\rm CubicSpline}(\log\kappa;\{p(\log\kappa_i)\}) ~\dd \log\kappa}\ ,
\end{equation*}
where ${\rm CubicSpline}(\log\kappa;\{p(\log\kappa_i)\})$ is the cubic spline interpolation function with given control point values $\{p(\log\kappa_i)\}$ (see red dots in \figref{fig:InferPkSpline} for the control points)
and we have applied normalization on the probability density function $p(\log\kappa)$ as in Eq.~(\ref{eq:norm}).
This population inference method is (almost) free of any assumption on the BNS mass function, the star formation rate and the delay time of BNS mergers thereafter,
therefore the reconstructed BNS population is expected to be unbiased.  We apply a uniform prior for the parameters, with the range of $p_i$ set as $[0.1 p_i^{\rm true}, 10 p_i^{\rm true}]$. The \texttt{DYNESTY} \cite{koposov2023Joshspeagle} sampler is used for the population inference, which implements the nested sampling algorithm \cite{skilling2006Nested}.

With this general population model, we constrain the population model parameters from the $N_{\rm det}$ detected events.
In \figref{fig:InferPkSpline}, we show the results of the BNS population inference $p(\log\kappa)$
for the two example cases $N_{\rm tot} =10^3, 10^4$. From \figref{fig:InferPkSpline}, we see that the probability 
$p(\log\kappa)$ is better constrained for high amplitudes  where the signals are louder
and is less constrained for low  amplitudes.
In the highest and the lowest $\kappa$ ends, we have fixed the probabilities as zero
as explained in the previous paragraph.

\begin{figure}
    \includegraphics[width=\columnwidth]{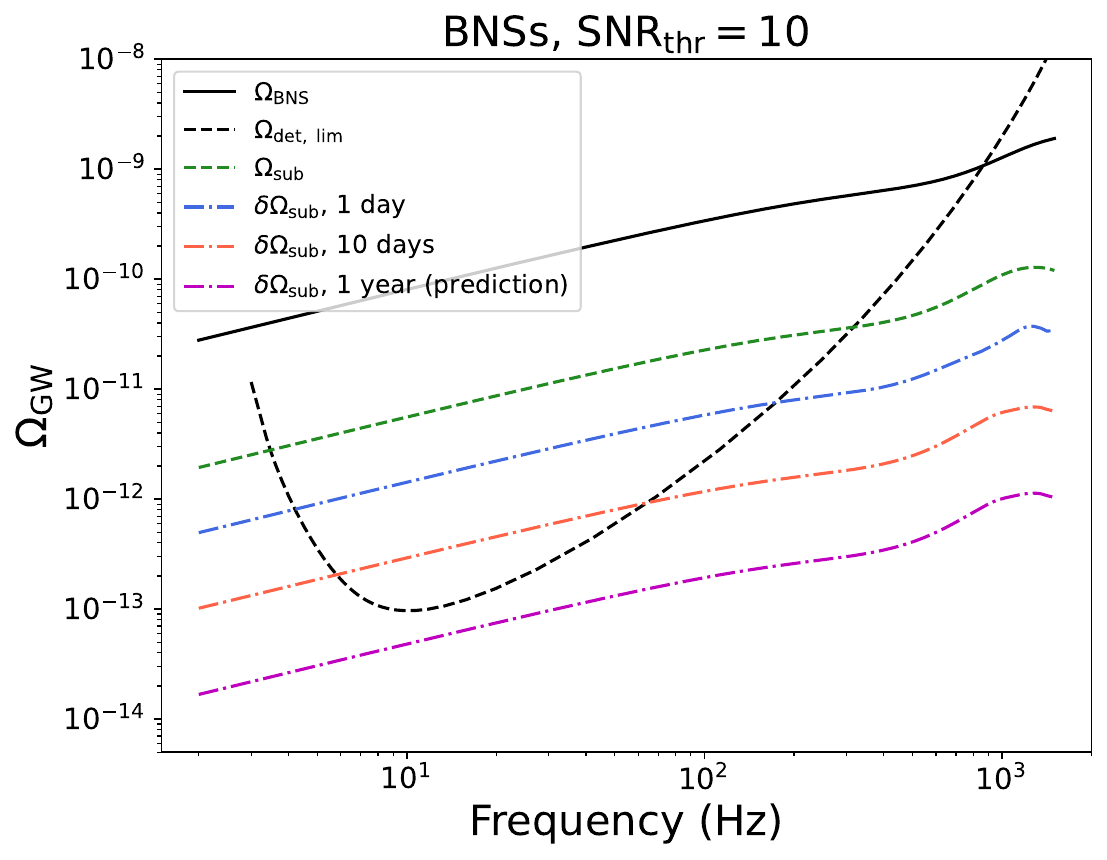}
    \caption{The energy density $\Omega_{\rm GW}(f)$ plot, where $\Omega_{\rm BNS}$ is the total foreground energy density from all BNSs, while $\Omega_{\rm sub}$ is contribution of subthreshold BNSs only, $\Omega_{\rm det.\ lim.}(f)$ is the detector sensitivity limit assuming a threshold ${\rm SNR}=3$ and a $T=1 ~{\rm yr}$'s observation period. 
    With $\sim 1$ day and $\sim 10$ days' observations, $\Omega_{\rm sub}$ is expected to be measured with fractional uncertainty $\delta\Omega_{\rm sub}/\Omega_{\rm sub}\approx 0.25$ and $\approx 0.05$, which will be largely reduced for a 1-year observation.}
    \label{fig:OmegaGWSpline}
\end{figure}

After the BNS population reconstructed, the energy density of the foreground from all the BNSs $\Omega_{\rm BNS}$ and subthreshold BNSs $\Omega_{\rm sub}$ can be calculated using \equref{eq:HA_sta}, where the selection effect is also taken into account in the latter calculations. We choose $68\%$ posterior samples from the constrained population model and calculate $\Omega_{{\rm sub},i}(f)$ for each sample population. The $1\sigma$ uncertainty range $\delta\Omega_{\rm sub}$ is therefore estimated as  $\delta\Omega_{\rm sub}(f)=\frac{1}{2}\left[\max_{i} \Omega_{{\rm sub},i}(f) - \min_{i} \Omega_{{\rm sub},i}(f)\right]$. As shown in \figref{fig:OmegaGWSpline}, in the fiducial population $\Omega_{\rm sub}(f)\approx  1.3\times 10^{-12} ~f^{2/3}_{\rm Hz}$, which is measured with fractional uncertainty $\delta\Omega_{\rm sub}/\Omega_{\rm sub} \approx 0.25$ from $\sim 1$ day's observations and $\delta\Omega_{\rm sub}/\Omega_{\rm sub} \approx 0.05$ from 10 days' observations. The fractional uncertainty of $\Omega_{\rm sub}$ roughly follows the scaling $\delta\Omega_{\rm sub}/\Omega_{\rm sub} \propto N_{\rm tot}^{-1/2}$. One can expect that the measurement precision would be largely improved with many more BNS events available in years long observations of 3G detectors, i.e.,
\begin{equation}
    \frac{\delta\Omega_{\rm sub}}{\Omega_{\rm sub}}= 0.05\left(\frac{10^4}{N_{\rm tot}} \right)^{1/2}=0.9\% \left(\frac{1\ {\rm yr}}{T} \right)^{1/2} \ ,
\end{equation}
where we have used the fiducial merger rate $\dot N$ [\equref{eq:MergerRateObs}] in the second equal sign.

For comparison, we also plot the detector sensitivity limit $\Omega_{\rm det. lim.}(f)$ which is the sensitivity of the detector network to the SGWB if no astrophysical foreground presented. We adopt a commonly used definition, the power-law integrated sensitivity curve proposed in \cite{Thrane2013}, assuming a threshold ${\rm SNR} =3$ and an observation time of $T=1 ~{\rm yr}$: any SGWB with energy density $\Omega_{\rm SGWB}(f)$ that is tangent to the detector sensitivity limit curve at $f_0$, i.e.,
\be
\Omega_{\rm SGWB}(f)= \Omega_{\rm det.lim.}(f_0)\times\left(\frac{f}{f_0}\right)^{\gamma_0}
\ee
with the power index $\gamma_0 = \frac{d\ln \Omega_{\rm det.lim.}(f_0)}{d\ln f_0}$ can be detected by the detector network with $3\sigma$ confidence level in 1 year (see \cite{Moore2015,Zhou2022} for the computational details).  

As shown in \figref{fig:OmegaGWSpline}, the measurement uncertainty $\delta\Omega_{\rm sub}(f)$ is below $\Omega_{\rm det.lim.}(f)$ in the entire frequency range. Therefore the sensitivity for detecting the primordial SGWB is mainly limited by the detector noise level and the total observation time, and the foreground of BNSs after proper cleaning is a minor limiting factor.

\subsection{Sanity check}

\begin{figure}
    \includegraphics[width=\columnwidth]{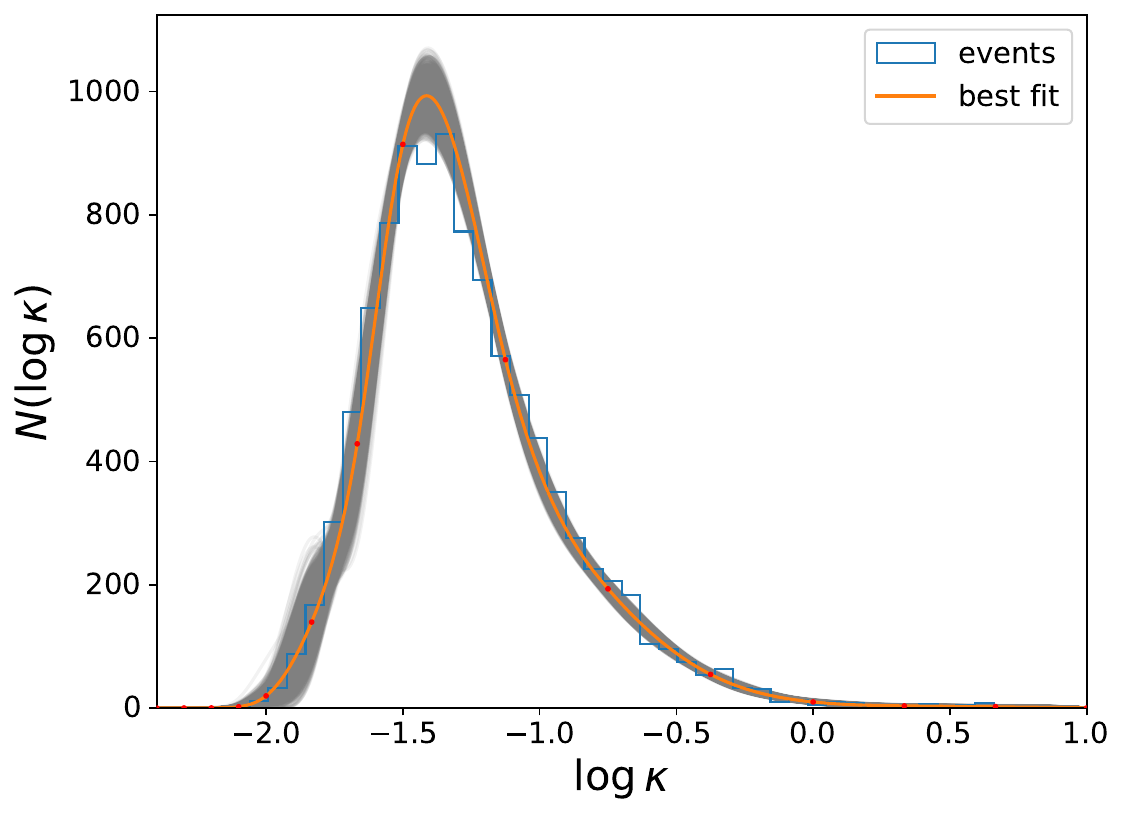}
    \caption{
    Constraint on the population model $p(\log\kappa)$ inferred from a total number of $N_{\rm tot} = 10^4$ BNSs with redshift dependent mass distribution $m_i\sim (1+z)^{0.1}$. }
    \label{fig:InferPkSplineZdep}
\end{figure}

As discussed in the previous subsection, 
this population inference method is (almost) free of any assumption on the BNS mass function, the star formation rate and the delay time of BNS mergers thereafter. As an example for demonstration, we inject a new population of BNSs with redshift dependent BNS mass distribution $m_i \sim (1+z)^\alpha$, considering possible dependence of neutron star masses on the cosmic metallicity. Specifically, we assume the local neutron star mass distribution is same to the one
considered in the previous subsection, $m_i\sim p_0(m_i)$.
The joint distribution of chirp mass and redshift is then formulated as 
\begin{equation}
    p(m_i,z) \propto p_0 \left(\frac{m_i}{(1+z)^\alpha} \right)\ .
\end{equation} 
We choose $\alpha=0.1$ and sample $N_{\rm tot}=10^4$ BNS merger events from the new population. The probability distribution $p(\log\kappa)$ is recovered using the same population inference method detailed in the previous two subsections, and the result is displayed in \figref{fig:InferPkSplineZdep}. It is of no surprise to find that the recovered population is unbiased. 
A closer comparison between \figref{fig:InferPkSplineZdep} and the right panel of \figref{fig:InferPkSpline} shows 
that the new population is slightly better constrained, simply because of louder signals in the new population.

\section{Summary and Discussion} \label{sec:Conclusion}

\subsection{Summary}
Foreground cleaning is essential for detecting the SGWB. 
In a nutshell, foreground cleaning is to measure and subtract the foreground energy density or equivalently the PSD.
There are in general two approaches in understanding and cleaning the foreground, an event-to-event approach in \equref{eq:HA_obs}, and a population based approach in \equref{eq:HA_sta}.
In the 3G era, most of BBHs and a fraction of BNSs are expected to be loud enough and can be individually detected. Their contribution to the foreground can be cleaned with high precision using the event-to-event approach, as shown in Paper I. However, cleaning the foreground of subthreshold BNSs has been a long-standing open question, which we aim to solve in this work.

We propose that the foreground PSD of subthreshold BNSs can be measured with the population based approach.
The basic idea is rather straightforward: the orientations of BNSs in the Universe should be isotropic, i.e., 
a uniform distribution in $\cos\iota$, therefore one can infer the number of subthreshold BNSs from individually detected ones and their contribution to the foreground PSD,
i.e., $p_{\rm sub}(\cos\iota|\kappa)+p_{\rm det}(\cos\iota|\kappa) = \mathcal{U}(-1,1)$.
In practice, the idea above is conducted in the framework of Bayesian population analysis.

As a convenient example, we constrain the BNS population model $p_{\rm pop}(\boldsymbol{\theta}|\mathbf{\Lambda})$ of BNS events sampled from an injection population $p_{\rm pop}^{\rm inj}(\boldsymbol{\theta}|\mathbf{\Lambda})$, with total number of events $N_{\rm tot}=10^3$ and $N_{\rm tot}=10^4$.  
With the constrained population model $p_{\rm pop}(\boldsymbol{\theta}|\mathbf{\Lambda})$ shown in \figref{fig:InferPkSpline}, we find the foreground energy density from subthreshold BNSs is measured with fractional uncertainty $\delta\Omega_{\rm sub}/\Omega_{\rm sub}\approx 0.25$ for $N_{\rm tot}=10^3$ and $\delta\Omega_{\rm sub}/\Omega_{\rm sub}\approx 0.05$ for $N_{\rm tot}=10^4$, where $\Omega_{\rm sub}(f) \approx 1.3\times 10^{-12} ~f^{2/3}_{\rm Hz}$. With a much higher number of BNS mergers available during years of observations of 3G detectors, a percent level measurement of $\Omega_{\rm sub}$ is expected (see \figref{fig:OmegaGWSpline}).

As a result, the residual foreground energy density from either the loud events or the subthreshold events after the foreground cleaning is 
expected to be below the detector sensitivity limit $\Omega_{\rm det.\ lim.}(f)$, therefore the astrophysical foregrounds from compact binaries will not be a limiting factor of the detection of the primordial SGWB with 3G detector network. Instead, it will be limited by the detector noise level and the total observation time, i.e., $\Omega_{\rm det.\ lim.}(f)$. 
Assuming the fiducial 3G detector network and a $1$-yr observation, a flat SGWB with $\Omega_{\rm SGWB}=10^{-13}$
is expected to be detected at 3 $\sigma$ confidence level  (see \figref{fig:OmegaGWSpline}).

\subsection{Discussions}

In this work, we have been focusing on cleaning the astrophysical foreground from subthreshold binary neutron stars. 
In fact, the cleaning methods proposed in paper I and in this work can be equally applied to other binaries, e.g.,
black hole-neutron star (BH-NS) binaries
(see e.g., Refs.~\cite{ET2023,Bellie:2023jlq} for current understanding of BH-NS merger rate and their contribution to the astrophysical foreground, where we can notice that the current understanding is still subject to large uncertainties).

In the population inference, we have been focusing on the major correlation between amplitude $\kappa$ and inclination angle $\iota$,
while neglecting their correlations with other parameters that are relatively weak as explained in the main text. 
In practice, when applying this foreground cleaning method to real data, the parameter samples of each detected event 
will be used in Eq.~(\ref{eq:popinfer}) where all the correlations among different parameters will be taken into account, strong or weak. 
Therefore, the predicted sensitivity might be mildly degraded by the weak correlations with other parameters.
Another possible limitation of this method is the assumption of 
low probability of BNS mergers at high redshifts ($z > 6$ in our fiducial model). This is a reasonable assumption 
according to the current understanding of star formation history and the BNS merger delay time \cite{KAGRA:2021duu}. 
But if this assumption breaks down due to some unknown physical processes, e.g., there was a large population 
of BNS mergers at high redshifts, say $z\approx 10$, 
then their contribution to the astrophysical foregrounds is beyond the 
horizon of the cleaning method proposed in this paper. In that case, we can only recover a subpopulation of BNSs and the inferred energy density $\Omega_{\rm sub}$ is a lower limit. Other methods, e.g., spectrogram correlated stacking proposed in \cite{Dey:2023oui}, maybe useful in this case.

Regarding the BNS mergers, we have assumed a rate that is consistent with previous LIGO/Virgo/KAGRA results. This fiducial merger rate might be an overestimate 
considering the nondetection of new BNS mergers in the ongoing run. Lower merger rate means the foreground cleaning problem is easier to deal with: using the fact that 
$\delta \Omega/\Omega \propto N_{\rm tot}^{-1/2}$ and $\Omega \propto N_{\rm tot}$, we find $\delta \Omega \propto \sqrt{N_{\rm tot}}$, i.e., the absolute magnitude of the residual energy density after foreground cleaning is lower.

The population based approach we proposed rely on the availability of parameter posteriors of individually detectable BNSs [Eq.~(\ref{eq:PopLikelihood})]. 
According to the estimate in a recent work \cite{Hu:2024mvn}, it takes millions of CPU hours for parameter inference of one-month catalog
with current accelerating techniques \cite{Green:2020dnx,Dax:2022pxd,Wong:2023lgb,Alvey:2023naa}. More efficient parameter estimation methods are necessary for cost-effective data analysis in the 3G era. With parameter posteriors of individually detectable BNSs, the hierarchical Bayesian inference can also be accelerated with neural network scheme \cite{Leyde:2023iof}.

We have adopted a commonly used detection threshold $\rho_{\rm thr}=10$. If we tune this threshold SNR to adjust the 
fractions of individually detected events and subthreshold events, a better threshold could be found for the purpose of  foreground cleaning and minimizing the residual energy density. This extension is straightforward to do, but does not add much to the cleaning method proposed in this work. Therefore, we decide not to investigate this extension here.

\acknowledgments
We thank Bei Zhou for sharing the detector sensitivity limit curve. This work was supported by the National Center for High-Level Talent Training in Mathematics, Physics, Chemistry, and Biology. This work made use of the Gravity Supercomputer at the Department of Astronomy, Shanghai Jiao Tong University.

\appendix

\bibliography{ms}
\end{document}